\documentclass[twocolumn,twocolappendix]{aastex631}

\newcommand{\hi}{\mbox{H\,{\sc i}}}
\newcommand{\logLx}{\log (L_\mathrm{X}/ \mathrm{erg\ s^{-1}})}
\newcommand{\logMs}{\log(M_\star/\mathrm{M_\odot})}
\usepackage{amsmath}
\newcommand{\vect}[1]{\boldsymbol{#1}}

\shorttitle{Nuclear X-ray of Antlia Member Galaxies}

\graphicspath{{./}{figures/}}

\usepackage{enumitem} 
\usepackage{amsmath}
\begin{document}

\title{AMUSE-Antlia I: NUCLEAR X-RAY PROPERTIES OF EARLY-TYPE GALAXIES IN A DYNAMICALLY YOUNG GALAXY CLUSTER}

\email{huzhensong@smail.nju.edu.cn, ysu262@g.uky.edu}

\author[0009-0009-9972-0756]{Zhensong Hu}
\affiliation{School of Astronomy and Space Science, Nanjing University, Nanjing $210023$, China}
\affiliation{Key Laboratory of Modern Astronomy and Astrophysics (Nanjing University), Ministry of Education, Nanjing 210023, China}
\author[0000-0002-3886-1258]{Yuanyuan Su}
\affiliation{Department of Physics and Astronomy, University of Kentucky, 505 Rose Street, Lexington, KY 40506, USA}
\author[0000-0003-0355-6437]{Zhiyuan Li}
\affiliation{School of Astronomy and Space Science, Nanjing University, Nanjing $210023$, China}
\affiliation{Key Laboratory of Modern Astronomy and Astrophysics (Nanjing University), Ministry of Education, Nanjing 210023, China}
\author[0000-0001-9662-9089]{Kelley M. Hess}
\affiliation{Department of Space, Earth and Environment, Chalmers University of Technology, Onsala Space Observatory, 43992 Onsala, Sweden}
\affiliation{Instituto de Astrof\'{i}sica de Andaluc\'{i}a (CSIC), Glorieta de la Astronom\'{i}a s/n, 18008 Granada, Spain}
\affiliation{ASTRON, the Netherlands Institute for Radio Astronomy, Postbus 2, 7990 AA, Dwingeloo, The Netherlands}
\author[0000-0002-0765-0511]{Ralph P. Kraft}
\affiliation{Harvard-Smithsonian Center for Astrophysics, 60 Garden Street, Cambridge, MA 02138, USA}
\author[0000-0002-9478-1682]{William R. Forman}
\affiliation{Harvard-Smithsonian Center for Astrophysics, 60 Garden Street, Cambridge, MA 02138, USA}
\author[0000-0003-0297-4493]{Paul E. J. Nulsen}
\affiliation{Harvard-Smithsonian Center for Astrophysics, 60 Garden Street, Cambridge, MA 02138, USA}
\affiliation{ICRAR, University of Western Australia, 35 Stirling Hwy, Crawley, WA 6009, Australia}
\author[0000-0002-7587-4779]{Sarrvesh S. Sridhar}
\affiliation{SKA Observatory, Jodrell Bank, Lower Withington, Macclesfield, Cheshire SK11 9FT, UK}
\author[0000-0001-8322-4162]{Andra Stroe}
\altaffiliation{Clay Fellow}
\affiliation{Harvard-Smithsonian Center for Astrophysics, 60 Garden Street, Cambridge, MA 02138, USA}
\author[0000-0002-3744-6714]{Junhyun Baek}
\affiliation{Department of Astronomy, Yonsei University, 50 Yonsei-ro, Seodaemun-gu, Seoul, 03722, Republic of Korea}
\author[0000-0003-1440-8552]{Aeree Chung}
\affiliation{Department of Astronomy, Yonsei University, 50 Yonsei-ro, Seodaemun-gu, Seoul, 03722, Republic of Korea}
\author[0000-0002-9961-3661]{Dirk Grupe }
\affiliation{Department of Earth and Space Sciences, Morehead State University, Morehead, KY 40514, USA}
\author[0000-0002-3612-9258]{Hao Chen}
\affiliation{Research Center for Intelligent Computing Platforms, Zhejiang Laboratory, Hangzhou 311100, China}
\author[0000-0003-4307-8521]{Jimmy A. Irwin}
\affiliation{Department of Physics and Astronomy, University of Alabama, Box 870324, Tuscaloosa, AL 35487, USA}
\author[0000-0003-2206-4243]{Christine Jones}
\affiliation{Harvard-Smithsonian Center for Astrophysics, 60 Garden Street, Cambridge, MA 02138, USA}
\author[0000-0002-3984-4337]{Scott W. Randall}
\affiliation{Harvard-Smithsonian Center for Astrophysics, 60 Garden Street, Cambridge, MA 02138, USA}
\author[0000-0003-2076-6065]{Elke Roediger}
\affiliation{E.A. Milne Centre for Astrophysics, School of Mathematics and Physical Sciences, University of Hull, Hull, HU6 7RX, UK}

\begin{abstract}
To understand the formation and growth of supermassive black holes (SMBHs) and their co-evolution with host galaxies, it is essential to know the impact of environment on the activity of active galactic nuclei (AGN).
We present new {\it Chandra} X-ray observations of nuclear emission from member galaxies in the Antlia cluster, the nearest 
non-cool core and the nearest merging galaxy cluster, residing at $D = 35.2\ \mathrm{Mpc}$.  Its inner region, centered on two dominant galaxies NGC~3268 and NGC~3258, has been mapped with three deep {\it Chandra} ACIS-I pointings.
Nuclear X-ray sources are detected in 
$7/84\ (8.3\%)$ early-type galaxies (ETG) and $2/8\ (25\%)$ late-type galaxies with a median detection limit of $8\times10^{38}\ \mathrm{erg\ s^{-1}}$.
All nuclear X-ray sources but one have a corresponding radio continuum source detected by MeerKAT at the L-band.
Nuclear X-ray sources detected in early-type galaxies are considered as the genuine X-ray counterpart of low-luminosity AGN.
When restricted to a detection limit of $\logLx\geqslant 38.9$ and a stellar mass of $10 \leqslant\log (M_\star/\mathrm{M_\odot})<11.6$, 6 of 11 ETG are found to contain an X-ray AGN in Antlia, exceeding the AGN occupation fraction of $7/39\ (18.0\%)$ and $2/12\ (16.7\%)$ in the more relaxed, cool core clusters, Virgo and Fornax, respectively, and rivaling that of the AMUSE-Field ETG of $27/49\ (55.1\%)$.
Furthermore, more than half of the X-ray AGN in Antlia are hosted by its younger subcluster, centered on NGC~3258.
We believe that this is because SMBH activity is enhanced in a dynamically young cluster compared to relatively relaxed clusters.
\end{abstract}

\keywords{black hole physics - galaxies: clusters: individual (Antlia)}

\section{Introduction}\label{sec:intro}

Nuclear X-ray emission provides an unambiguous diagnostic of the activity of supermassive black holes (SMBH). Its correlation with black hole mass and its host galaxy is fundamental to understanding the relation between SMBH and the host galaxy properties, such as the $M_\mathrm{BH}$--$\sigma$ and $M_\mathrm{BH}$--$L_\mathrm{X}$ relations \citep{2013ARA&A..51..511K,2019ApJ...884..169G}. The sub-arcsec spatial resolution of {\it Chandra} has made it possible to detect nuclear X-ray emission down to $\sim10^{38}\ \mathrm{erg\ s^{-1}}$, enabling the study of active galactic nuclei (AGN) in a large number of low mass quiescent galaxies.  The AGN Multi-wavelength Survey of Early-Type Galaxies in the Virgo Cluster (AMUSE-Virgo) has studied the nuclear X-ray emission of $100$ elliptical, lenticular, and dwarf elliptical galaxies in Virgo, a well known cool core cluster with a sharply increasing X-ray surface brightness profile towards its center. 
It was found that $24\%-34\%$ of the Virgo early-type galaxies (ETG) host X-ray AGN. Also, it provided evidence for down-sizing: black holes with lower mass radiate closer to their Eddington limits than their higher mass counterparts \citep{2008ApJ...680..154G,2010ApJ...714...25G}. A study of another cool core cluster, Fornax \citep{2019ApJ...874...77L}, reports a level of nuclear activity similar to Virgo, with $27\%\pm10\%$ ETG hosting AGN.
 
 The environment plays a key role in galaxy evolution. Relaxed cool core clusters are dominated by red, elliptical galaxies, due to a number of quenching mechanisms, including ram pressure stripping \citep[e.g.][]{1972ApJ...176....1G,1980ApJ...236..351D}. The dependence of the nuclear X-ray activity on the large scale environment can provide insight into the mechanisms that govern the feeding and feedback of SMBH. AMUSE-Field is a {\it Chandra} large program 
 targeted on $103$ nearby field and group ETG for a comparison with AMUSE-Virgo. In this paper, ``the Field" stands for the AMUSE-Field, which refers to heterogeneous environments from galaxy groups to isolated fields. \citet{2012ApJ...747...57M} report that the AMUSE-Field sample displays a higher X-ray AGN occupation fraction $45\%\pm7\%$ and a higher nuclear X-ray luminosity at a given black hole mass than the Virgo sample. \cite{2019ApJ...874...77L} has further confirmed that AMUSE-Field is also more active than Fornax. Environments like the Virgo and Fornax clusters may have suppressed black hole accretion and quenched star formation by cutting off the fuel supply via ram pressure stripping \citep{2020ApJ...895L...8R}. However, the intrinsic properties of various galaxy clusters can be different. For example, member galaxies in dynamically young, merging clusters as well as high redshift proto-clusters, often have enhanced star formation and contain abundant cold gas comparable to field galaxies \citep{2017MNRAS.465.2916S,2017A&A...606A.108C,2017ApJ...842L..21N}.
 The difference of black hole activity in a variety of nearby clusters can cast light on the environmental dependence of black hole activity.

The Antlia cluster (Abell S0636) is the third nearest cluster after Virgo and Fornax at a distance of $35.2\ \mathrm{Mpc}$ ($1''=170\ \mathrm{pc}$) \citep{2003A&A...408..929D}. It is a Bautz-Morgan type III cluster.  Antlia has a $R_\mathrm{200}$\footnote{$R_\Delta$ is the radius within which the enclosed matter density is $\Delta$ times the critical density of the universe. $R_{200}$ is conventionally taken as an approximate of the virial radius of a cluster.} of $887\ \mathrm{kpc}$ and $M_\mathrm{200}$ of $7.9\times10^{13}\ \mathrm{M_\odot}$\citep{2016ApJ...829...49W}. Its size and halo mass are similar to the Virgo cluster with $R_\mathrm{200} = 974.1\pm5.7\ \mathrm{kpc}$ and $M_\mathrm{200}=1.05\pm0.02\times10^{14}\ \mathrm{M_\odot}$\citep{2017MNRAS.469.1476S} and the Fornax cluster with $R_\mathrm{200} \sim 700\ \mathrm{kpc}$ and $M_\mathrm{200}\sim7\times10^{13}\ \mathrm{M_\odot}$ \citep{2001MNRAS.326.1076D}.  Meanwhile, the global temperature of Antlia is $2\ \mathrm{keV}$ \citep{2016ApJ...829...49W}, which falls between that of Virgo of $2.3\ \mathrm{keV}$ \citep[e.g.][]{2011MNRAS.414.2101U} and Fornax of $<1.5\ \mathrm{keV}$ \citep[e.g.][]{2017ApJ...851...69S,1997ApJ...482..143J}. 
Antlia is likely the dynamically youngest of these three galaxy clusters.
The main cluster of Antlia, centered on the brightest cluster galaxy NGC~3268, is in the process of merging with a subcluster associated with the bright elliptical galaxy NGC $3258$, which is $22'$ ($225\ \mathrm{kpc}$) to the southwest of NGC $3268$.
\textit{ASCA}, \textit{Suzaku}, and \textit{XMM-Newton} observations have revealed that its ICM displays relatively uniform surface brightness and temperature distributions at the cluster center, in contrast to typical cool core clusters with a sharp surface brightness peak and a steep temperature gradient \citep{2000PASJ...52..623N,2016ApJ...829...49W}.
 The galaxy density of Antlia is $1.7$ times higher than Virgo and $1.4$ times higher than Fornax \citep{1990AJ....100....1F}. Also, the ratio of the velocity dispersion between infalling galaxies to virialized galaxies $\sigma_{\rm infall}/\sigma_{\rm vir}$ in Antlia is $2.31$ \citep{2015MNRAS.452.1617H}, higher than that of Virgo of $1.64$ and the predicted virialized population ratio of $1.4$ \citep{2001ApJ...559..791C}, suggesting that the Antlia cluster is less virialized. CO ($2\textup{--}1$) and \hi\ observations reveal that many of Antlia's member galaxies, both star forming and passive, contain large reservoirs of molecular and atomic gas, unlike galaxies in more relaxed clusters \citep{2015MNRAS.452.1617H,2019ApJ...882..132C}.

To study the nuclear X-ray activity in Antlia, 
we observed the central region with three \textit{Chandra} ACIS-I pointings (PI: Y. Su), with an exposure of $\sim 70\ \mathrm{ks}$ each and $223.9\ \mathrm{ks}$ in total.
As shown in Figure \ref{fig:DSS}, the three AMUSE-Antlia fields focused on: NGC $3268$, NGC $3258$, and the southeast of NGC $3268$.  
There are $92$ galaxies covered by observations that have stellar masses $M_\star$ in the range of $10^{7}\textup{--}10^{11}\ \mathrm{M_\odot}$, 84 of which are ETG while 8 are late-type galaxies (LTG). 

 This paper is structured as follows. 
Data preparation and methods are presented in Section \ref{sec:data_preparation}. The results of the nuclear X-ray source detection are shown in Section \ref{sec:result}. Section \ref{sec:ETG_AGN_diffSamples} compares the AGN occupation fraction and X-ray luminosity function (XLF) of ETG in Antlia with AMUSE-Virgo, AMUSE-Field, and Fornax ETG AGN.
Our findings are discussed in Section~\ref{sec:discussion} and summarized in Section~\ref{sec:conclusion}.

\section{Data Preparation}\label{sec:data_preparation}

\begin{figure*}[ht!]
\epsscale{0.8}
\plotone{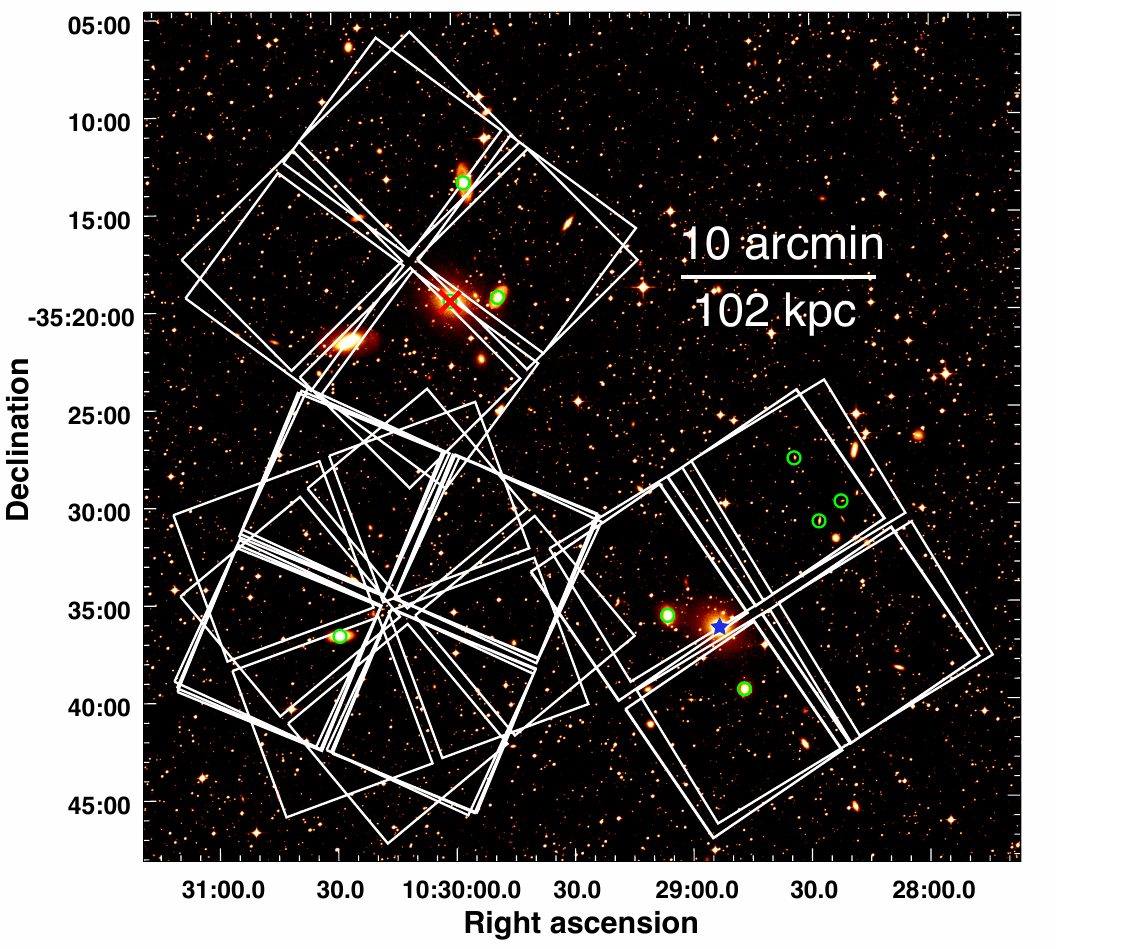}
\caption{DSS $J$ band image of the central region of the Antlia cluster, $10' \simeq 102\ \mathrm{kpc}$. The fields of the {\it Chandra} observations presented in this study are marked in white. 
Green circles indicate the detected nuclear X-ray sources. NGC $ 3268 $ and NGC $ 3258 $ are marked with a red ``$\times$" and a blue ``$\star$", respectively.   \label{fig:DSS}}
\end{figure*}

We reprocessed the \textit{Chandra} level-1 data and the calibration files according to the standard procedure of CIAO v$4.14$ \citep{2006SPIE.6270E..1VF}.  To calibrate the astrometry of each field, we chose the longest-exposed image as the reference image and matched the centroid of commonly detected point sources with the CIAO tool {\tt reproject\_aspect}. We checked the lightcurves for flares. Counts maps, exposure maps, and point spread function (PSF) maps were generated for each observation in three energy bands:  $0.5$--$2$ ($S$-band), $2$--$8$ ($H$-band), and $0.5$--$8$ ($F$-band) $\mathrm{keV}$. The maps of multiple observations of the same field were then merged.

We followed the source detection procedures described in \citet{2017ApJ...846..126H} and \citet{2019ApJ...876...53J}. The original X-ray point source list was generated by the CIAO tool {\tt wavdetect}. To correct for the source centroids, we iterated over the source position within the $90\% $ PSF. The position uncertainty (PU) at $68\%$ confidence level was calculated according to the empirical relation among PU, off-axis angle (OAA, unit in arcminutes), and source counts ($C$) \citep[Equation 14]{2007ApJ...659...29K},

\begin{equation}\small
\begin{array}{l}
\log \text{PU} = \\
0.114 \text{OAA}-0.460 \log C - 0.240, 0.000<\log C \le2.123,  \\
 0.103\text{OAA}-0.195\log C - 0.803,  2.123<\log C \le 3.300.
\end{array}
\end{equation}

To filter out spurious sources due to background fluctuations, we calculated the binomial no-source probability $P_\mathrm{B}$ \citep{2007ApJ...657.1026W} and removed those sources with $P_\mathrm{B} > 0.01$. Finally, a cross-matching method \citep{2009ApJ...706..223H} was applied to identify the same source detected in different energy bands. We only kept X-ray sources located within $8'$ from the aimpoint to ensure that they are covered by all observations with the same aimpoint but different roll angles. The resultant source catalog and a more detailed data processing and source detection method will be fully presented in a separate publication (in preparation).

To identify AGN, we searched for optical nuclei coincident with any X-ray point-like source emission. The AMUSE-Antlia footprint contains 92 member galaxies, according to the member galaxy catalog based on the optical observation using the $4$-m Blanco telescope at CTIO \citep{2020MNRAS.497.1791C}. We calculated the stellar masses using the mass-luminosity relation \citep{2003ApJS..149..289B}. We measured the $g-r$ band color index and $r$ band luminosity $L_r$ of each galaxy from the CTIO image. The stellar mass was then determined as $\log (M/L) = -0.306+1.097\ (g-r)$. We compared the galaxy positions with those in the 2MASS Extended Source Catalog (XSC) \citep{2006AJ....131.1163S}. We adopted the XSC coordinates if the galaxy falls within $1''$ of any XSC records. Then we identified AGN by looking for any X-ray point source that is located within the minimum of $1''$ or $3$ PU from the optical nuclei of any member galaxy.  
The Digital Sky Survey (DSS) $J$ band image of the Antlia central region is shown in Figure \ref{fig:DSS}, where $10' \simeq 102\ \mathrm{kpc}$. The field of view (FoV) of {\it Chandra} ACIS-I observations presented in this study are highlighted with white solid boxes. The positions of detected nuclear X-ray point sources are marked as green circles.

In addition, we compared the nuclear X-ray sources with their corresponding MeerKAT radio continuum image (see Figure \ref{fig:NGalaxy_p2}). The MeerKAT observations (SCI-20210212-KH-01; PI: K. Hess) were carried out in the L band, spanning the 856\textup{--}1712 MHz frequency range and covering a region out to 1.4 times the virial radius of the cluster.  A forthcoming publication will present a detailed analysis of the MeerKAT spectropolarimetry data. The radio continuum image has an rms noise of about 6.5 $\mu$Jy/beam and an angular resolution of $7''$ with an astrometric uncertainty of $1.5''$.


\section{Results}\label{sec:result}
\subsection{Nuclear X-ray Emission}
We find $ 9 $ point-like X-ray sources located at the optical nuclei of the member galaxies. As discussed in \citet{2008ApJ...680..154G}, these sources are generally considered AGN, although some of them could be X-ray binaries (XRB).  $7$  of the $9$ host galaxies are ETG, while the other $2$ are LTG -- a blue compact dwarf (BCD), Antlia $98$, and a spiral galaxy, Antlia $88$.  

\begin{figure*}[ht!]
\epsscale{1.15}
\plotone{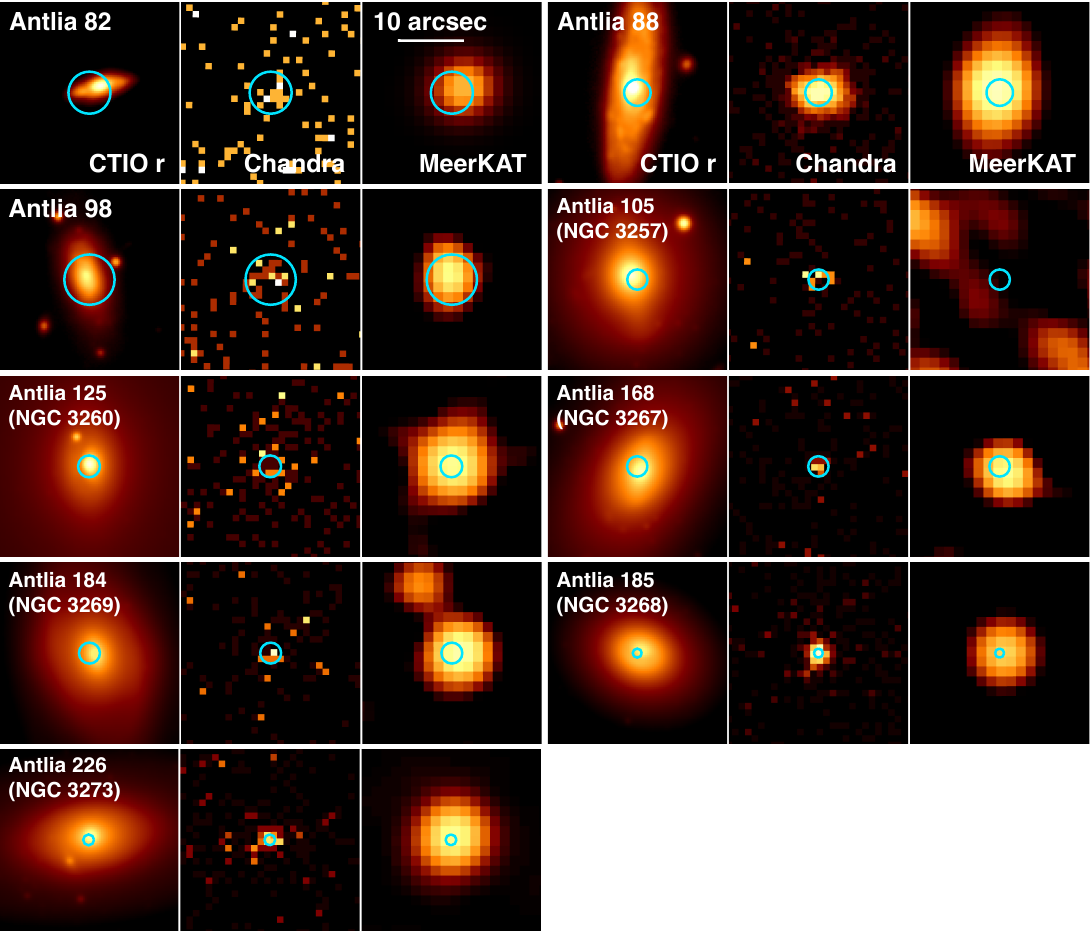}
\caption{Multi-wavelength images of the galaxies that host nuclear X-ray sources, $10''\simeq1.7\ \mathrm{kpc}$. For each galaxy, from left to right, are images of CTIO $r$ band, {\it Chandra} X-ray in the energy band of $0.5$--$8\ \mathrm{keV}$, and broad bandwidth L-band radio continuum image from MeerKAT. The X-ray image is binned into twice the original pixel size. The name of each galaxy at the upper right corner of the CTIO $r$ image is adopted from \citet{1990AJ....100....1F}. The NGC name is also shown if available. The cyan circle centered on each X-ray source has a radius of $50\%$ PSF. All galaxies have a corresponding radio source except Antlia $105$.}  
 \label{fig:NGalaxy_p2}
\end{figure*}

 The galaxy and nuclear X-ray source properties are summarized in Table \ref{table:AGN}. The CTIO $r$ band, {\it Chandra} X-ray,  and MeerKAT radio images of each galaxy in which a nuclear X-ray source is detected are shown in Figure \ref{fig:NGalaxy_p2}. The estimated black hole masses range from $3\times10^5\ \mathrm{M_\odot}$ to $5\times10^7\ \mathrm{M_\odot}$, with an uncertainty of $0.3\textup{--}0.4$ dex, according to the fundamental plane (Appendix \ref{appendix:FundPlane}).

No nuclear X-ray source is detected in NGC $3258$, the dominant elliptical galaxy of the southern subgroup that is merging with the northern group centered on NGC~3268. The X-ray emission at NGC $3258$ is quite extended, which can lead to spurious detection\footnote{At the first stage of the X-ray source detection process, there are two X-ray point sources ``detected'' close to the center of NGC $3258$: one is in the $S$-band that offsets by $0.77''$ from the center, the other is in the $H$ and $F$-band and offsets by $1.66''$. These two sources are more likely to be false detections caused by the diffuse emission.}. We fit the spectrum of the central region to an absorbed power-law model and obtained a power-law index of about $ 4$, which is about twice the typical spectral index of an AGN\footnote{The spectra can also be fitted with an absorbed thermal plasma model ({\tt apec} in XSPEC), with a galactic absorption of $N_\mathrm{H} \approx 2.8\times10^{21}\ \mathrm{cm^{-2}}$ and a plasma temperature of $0.8\ \mathrm{keV}$.}. It is not considered an AGN due to its extended shape and soft spectrum.

The AGN detection in ETG can be contaminated by low mass X-ray binaries (LMXBs). The distribution of LMXB follows the stellar mass distribution, which can be traced by a S\'ersic profile \citep{1968adga.book.....S}. The projected stellar mass inside the $1''$ nuclear source matching radius accounts for $\sim 3\%$ of the total, assuming a mean effective radius of $\sim 9''$. We adopt the LMXB X-ray luminosity function \citep{2012A&A...546A..36Z} to estimate the number of LMXBs over the detection limit. 
We expect $\sim 0.4$ X-ray nuclear sources to be LXMB, which is small compared to the 7 X-ray nuclear sources
detected in ETG in AMUSE-Antlia.

\begin{deluxetable}{ccccccc}
\tablenum{1}
\tablecaption{Nuclear X-ray source and host galaxy properties}
\tablewidth{0pt}
\tabletypesize{\footnotesize}
\tablehead{
\colhead{FS$90$} & \colhead{ NGC }& RA & DEC & \colhead{Morph. } &\colhead{$\log L_\mathrm{X}$}  & \colhead{$\log M_\star$} \\
& &   ($\mathrm{deg}$) &   ($\mathrm{deg}$) &  & ($\mathrm{erg\ s^{-1}}$) & ($\mathrm{M_\odot}$) 
}
\decimalcolnumbers
\startdata
82 & & 157.09246	&  -35.49791 &  E & $38.86^{+0.20}_{-0.15}$ & $9.2$ \\
88 & &157.11674 &	-35.51491 & L & $41.53^{+0.01}_{-0.01}$ & $9.3$ \\
 98 & &157.14249  & 	-35.46079  &L & $39.05^{+0.16}_{-0.12}$ & $8.7$ \\
105 & 3257 & 157.19617 &	-35.65797  & E & $38.92^{+0.12}_{-0.10}$ & $10.4$ \\
125 & 3260 & 157.27638 &	-35.59513 & E & $39.03^{+0.11}_{-0.10}$ & $10.6$ \\
168 & 3267 & 157.45242 &-35.32195 & E & $39.07^{+0.10}_{-0.09}$ & $10.5$ \\
184 & 3269  & 157.48765 &	-35.22433& E & $39.14^{+0.11}_{-0.10}$ & $10.8$ \\
185 & 3268 & 157.50272 &	-35.32545 & E & $40.75^{+0.02}_{-0.02}$ & $11.6$ \\
226 & 3273 & 157.62125 &	-35.61017 & E & $39.27^{+0.08}_{-0.07}$ & $10.8$
\enddata
\tablecomments{($ 1 $) Galaxy Name according to \citet{1990AJ....100....1F}. (2)  NGC name of the galaxy.  (3\textup{--}4) Right ascension and declination at equinox J2000. ($5$) The morphological type of the galaxy, ``E" and ``L" stand for ETG and LTG.  ($6$) X-ray luminosity in the $0.5$--$8\ \mathrm{keV}$ energy  band. ($ 7 $) Stellar mass of the galaxy, derived from the $g-r$ mass-luminosity relation, with an uncertainty of $\sim 0.2$ dex.}
\label{table:AGN}
\end{deluxetable}

\subsection{The nuclear X-ray source in a BCD} \label{ssec:bcd_xrc}
Blue compact dwarf galaxies resemble galaxies in the infant universe, a critical stage for the formation of black holes and the establishment of the M--$\sigma$ relation. 
An X-ray and a radio source are found overlapping with one BCD, Antlia 98. The membership of the galaxy to the Antlia cluster is confirmed by \citet{2008MNRAS.386.2311S}. It is classified as a BCD based on optical and $\mathrm{H\alpha}$ observations \citep{1990AJ....100....1F,2008MNRAS.386.2311S,2014A&A...563A.118V}.  \citet{2014A&A...563A.118V} find two non-central star forming knots in Antlia 98. 
The spatial relationship between the X-ray source and the two star forming knots is unclear, as
this source is close to the edge of the FoV, where the $50\%$  PSF is sizable, with a positional uncertainty of $4''$. The directly measured photon flux is $F_\mathrm{0.5\textup{--}8\ keV} = (2.2\pm0.7)\times10^{-6}\ \mathrm{ph\ cm^{-2}\ s^{-1} }$, which corresponds to an X-ray luminosity of $L_\mathrm{0.5\textup{--}8\ keV}  = (1.1\pm 0.3)\ \times\ 10^{39}\ \mathrm{erg\ s^{-1}}$, and an H-band luminosity of $L_\mathrm{2\textup{--}10\ keV}  = (7.7 \pm 2.3)\ \times\ 10^{38}\ \mathrm{erg\ s^{-1}}$, assuming a power-law photon index of $1.8$ and a galactic absorption of $N_\mathrm{H} = 1\times10^{21}\ \mathrm{cm^{-2}}$. 
The $1.28\ \mathrm{GHz}$ flux measured from a MeerKAT observation is $(8.4\pm 1.5 )\times10^{-5}\ \mathrm{Jy}$, corresponding to a star formation rate of $\mathrm{SFR} = (7.8\pm 1.3) \times10^{-3}\ \mathrm{M_\odot\ yr^{-1}}$ \citep{2012ARA&A..50..531K}. Considering the galaxy stellar mass of $ M_\star = (5.2\pm1.2) \times 10^{8}\ \mathrm{M_\odot}$, the specific star formation rate is $\mathrm{SFR} / M_\star = (1.5\pm 0.2)\times10^{-11}\ \mathrm{yr^{-1}}$, which is relatively small compared to other BCDs \citep{2010AJ....139..447H}.

We investigate whether this source is an XRB or an SMBH. \citet{2010ApJ...724..559L} notice a tight correlation among the $2\textup{--}10\ \mathrm{keV}$ hard band X-ray luminosity of XRB, stellar mass, and SFR, which is $L_{2\textup{--}10\ \mathrm{keV}} = \alpha M_\star + \beta \mathrm{SFR}$, where $\alpha = (9.05\pm0.37)\times10^{28}\ \mathrm{erg \ s^{-1}\ M_\odot^{-1}}$ and $\beta = (1.62\pm0.22)\times10^{39}\ \mathrm{erg \ s^{-1}\ M^{-1}_\odot\ yr}$. For Antlia 98, we obtain a $2\textup{--}10\ \mathrm{keV}$ X-ray luminosity of $(6.0\pm 0.3)\times10^{37}\ \mathrm{erg\ s^{-1}}$ for the expected XRB dominated X-ray emission. This falls short of the detected X-ray luminosity but consistent within the $3\sigma$ uncertainty. 
The ratio of the X-ray and radio intensities can also cast light into its origin. We follow the correlation given by \citet{2003ApJ...583..145T},

\begin{equation}
R_{\rm X}=\frac{\nu L_{\nu}(\rm 5\, GHz)}{L_{\rm X}\rm (2-10)\,keV}\ .
\end{equation}
We derive the source 5\,GHz luminosity of  $ (6.1\pm1.1)\times10^{34}\  \mathrm{ erg\ s^{-1}}$ from its 1.28\,GHz luminosity based on an assumed power-law spectrum
$S\propto \nu^{-0.7}$ \citep{2002AJ....124..675C}. We obtain log\,$R_{\rm X}\sim-4.1$. This source is too luminous in the radio to be a stellar-mass XRB, which typically have log\,$R_{\rm X}\leq-5.3$. Therefore, the source detected at the center of Antlia 98 is a promising AGN candidate. Future on-axis observation is required to determine its nature.

\subsection{X-ray stacking of undetected galaxies}
To probe the AGN population below the detection limit, we performed a stacking analysis for member galaxies lacking X-ray detected AGN. Among the $83$ candidate galaxies, we exclude NGC $3258$ due to the diffuse nature of its X-ray emission. Based on the PSF and stellar mass $M_\star$, we categorize the galaxies into four groups: small-PSF low-mass subset (SPLM), small-PSF intermediate-mass subset (SPIM), intermediate-PSF low-mass subset (IPLM), and intermediate-PSF intermediate-mass subset (IPIM). The boundary of small and intermediate $90\%$ PSF is $4''$, while that of the low and intermediate mass is $ \logMs= \ 8.5$.  There are $20$, $11$, $30$, and $21$ galaxies in the SPLM, SPIM, IPLM, and IPIM subsets, respectively. We stack the counts maps of galaxies in each subset, and extract the net counts within a nuclear region $2''$ in radius, for which we chose an annulus with an inner radius of $4''$ and an outer radius of $5''$ as the background. We use the CIAO tool {\tt aprates} to compute the net counts and uncertainties for each subset. No signal is detected in the LM subset for either PSF. Taking $N_\mathrm{H} = 1\times10^{21}\ \mathrm{cm^{-2}}$ and $\Gamma = 1.8$, the $3\sigma$ upper limits on the unabsorbed $0.5 - 8\ \mathrm{keV}$ luminosity for SPLM and IPLM are $3.3\times10^{37}\ \mathrm{erg\ s^{-1}}$ and $3.0\times10^{37}\ \mathrm{erg\ s^{-1}}$, respectively. However, X-ray emission is detected for the subsets of more massive galaxies. A detection with a signal-to-noise ratio (SNR) of $3.1$ is obtained for SPIM. There are $22$ net counts, with $41$ and $44$ counts in the source and background apertures, respectively, corresponding to a photon flux of  $(1.7\pm 0.6) \times10^{-7}\ \mathrm{ph\ cm^{-2}\ s^{-1}}$, and an  unabsorbed luminosity of $(8.0\pm2.6)\times10^{37}\ \mathrm{erg\ s^{-1}}$. For the IPIM subset, we obtain a SNR of $2.1$ and an unabsorbed luminosity of $(4.0\pm1.8)\times10^{37}\ \mathrm{erg\ s^{-1}}$. 
Fainter nuclear X-ray sources are likely missed due to the limited sensitivity.

\section{Black hole activity in ETGs}\label{sec:ETG_AGN_diffSamples}

\subsection{Nuclear X-ray Luminosity Function}\label{sec:XLF}

\begin{figure*}[ht!]
\epsscale{1}
\plottwo{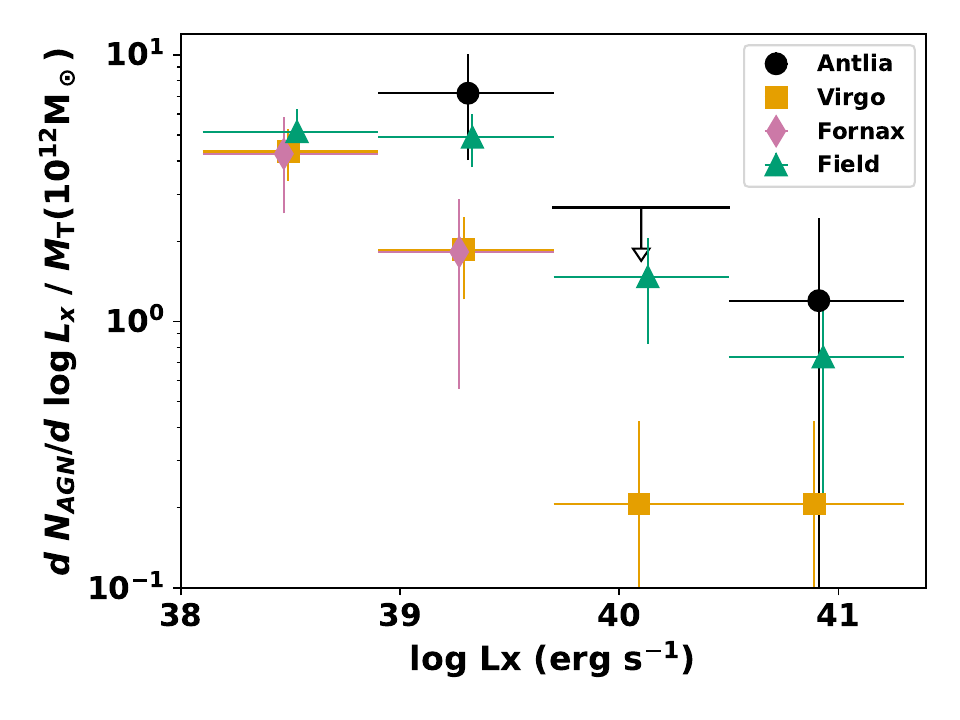}{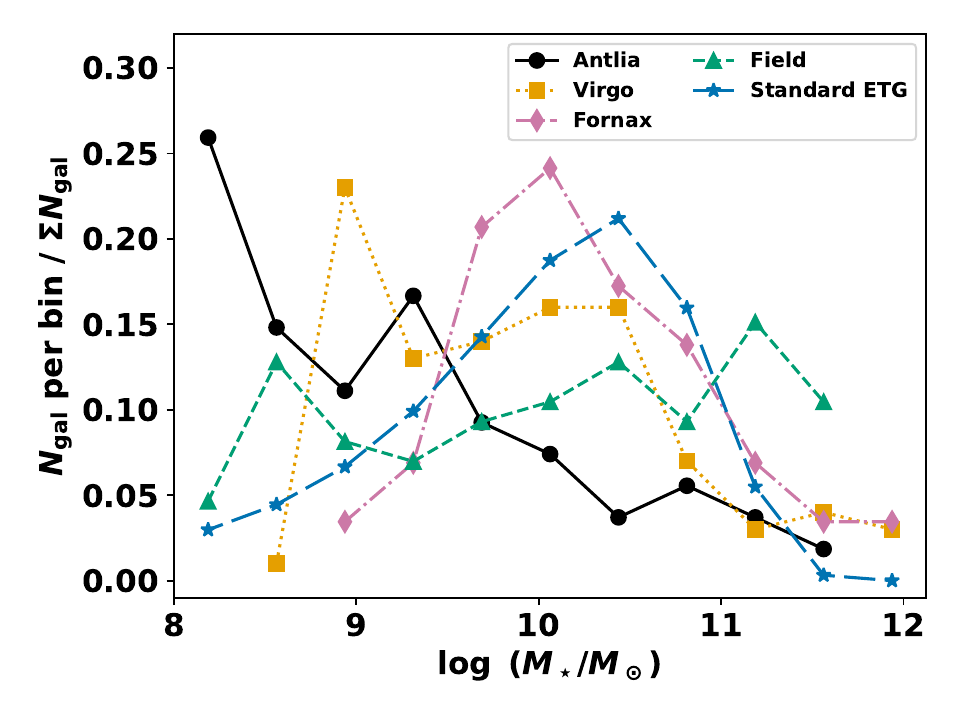}
\caption{Left panel: ETG XLF for AMUSE-Antlia (black error bar), AMUSE-Field  (green error bar), AMUSE-Virgo (yellow error bar), and Fornax sample (pink error bar). The leftmost luminosity bin is under the median sensitivity of AMUSE-Antlia, so we focus on the right three bins. We present the $90\%$ upper limit as a downwards arrow for the non-detection of the third bin of Antlia XLF.  The $y-$axis is log$L_\mathrm{X}$ number density divided by the total stellar mass of galaxies $M_\mathrm{T}$  in each sample.  Right panel: The stellar mass distributions of the four samples. Also, we plot the standard ETG stellar mass function \citep{2016MNRAS.457.1308M} with the blue long-dashed line. Below a certain detection limit, high mass galaxies can have more AGN detections due to Eddington incompleteness, so for the left panel we divide the XLF by $M_\mathrm{T}$. Generally speaking, the AGN XLF in Antlia is consistent with the Field, while both are much higher than Virgo and Fornax. \label{fig:AGN_XLF}}
\end{figure*}

We compare the ETG AGN X-ray Luminosity Functions (XLF) of the Antlia cluster, as well as the Virgo cluster \citep{2010ApJ...714...25G} , the Fornax cluster \citep{2019ApJ...874...77L}, and the AMUSE-Field \citep{2012ApJ...747...57M} in Figure \ref{fig:AGN_XLF}.  The detection limits of these surveys are different. The median and completeness sensitivities of AMUSE-Antlia footprint are $\logLx = 38.9$ and $\logLx = 39.0$ in the $0.5$\textup{--}$8\ \mathrm{keV}$ passband, which are the shallowestamong all the samples.  The AMUSE-Field, AMUSE-Virgo, and Fornax survey are dominated by snapshots, which means that the target galaxies are at the aimpoint.  For the snapshots, the detection limit 
mainly depends on the effective exposure time, while in the AMUSE-Antlia fields, the detection limit also depends on the off-axis angle. The completeness sensitivities for AMUSE-Field, AMUSE-Virgo, and Fornax snapshots are $38.3, 38.6$, and $38.7\ \mathrm{dex}$, respectively.
To compare all AGN presented in those studies, while taking into account the sensitivity difference, we set the median sensitivity of AMUSE-Antlia at the beginning of the second bin. Thus, all the samples are comparable, except for the faintest bin.

AGN detection can be biased by the ``Eddington ratio incompleteness'' \citep{2010ApJ...714...25G,2012ApJ...747...57M}. For example, \citet{2010ApJ...714...25G} emphasize that the nuclear SMBH activity does not increase with the host stellar mass, if the sample is Eddington complete. On the other hand, for any luminosity-limited survey, it is impossible to reach the same Eddington-scaled luminosity across an extensive range of black hole masses. It is therefore more likely to detect SMBH activity in more massive galaxies, due to their higher luminosity and possibly higher black hole masses. The stellar mass distributions of the four samples are different, as shown in the right panel of Figure \ref{fig:AGN_XLF}. There are more low mass galaxies in the Antlia sample than any other, while the AMUSE-Field sample contains more high mass galaxies.  We apply the Kolmogorov–Smirnov test (K-S test) to examine if any two samples are from the same distribution. The results of the K-S test indicate significant differences in the stellar mass distributions between each pair of samples. The p-values for the comparisons between the sample pairs Antlia and Virgo,  Antlia and Field, Antlia and Fornax,  Virgo and Field, Virgo and Fornax, and Field and Fornax are $1.1\times10^{-16}$, $2.5\times10^{-6}$, $1.2\times10^{-11}$,  $1.42\times10^{-4}$, $7.4\times10^{-3}$, and $1.7\times10^{-3}$, respectively.  Also, we compare these samples with the standard ETG stellar mass distribution \citep{2016MNRAS.457.1308M} of the Galaxy and Mass Assembly survey phase two (GAMA-II) survey down to a completeness limit of $\logMs = 8$. The p-value for Fornax is $0.33$ and therefore the Fornax sample is consistent with the standard ETG distribution. Other samples deviate significantly from the standard distribution.

To compare samples with different stellar mass distributions, we divide the number of galaxies by the total stellar mass of all ETG in each sample, the total stellar masses $M_\mathrm{T}$ for Antlia, Field, Virgo, and Fornax are $1.0$, $5.1$, $6.0$, and $2.1$, in units of $10^{12}\ \mathrm{M_\odot}$. The intrinsic X-ray luminosities are converted to $0.5$--$8\ \mathrm{keV}$ assuming a power-law photon index $\Gamma=1.8$, the median value of AGN spectra in the local $50\ \mathrm{Mpc}$ volume \citep{2017ApJ...835..223S}. Also, we apply the same luminosity bins to all data. The lowest luminosity bin is lower than the median sensitivity of most samples, so we focus on the other three bins. No Fornax ETG is detected in the two most luminous bins. No Antlia ETG AGN is detected at $\logLx =39.7 \textup{--} 40.5$, while only one AGN with $\logLx=40.75\pm0.02$ falls in the $\logLx =40.5 \textup{--} 41.1$ bin. The high luminosity bin is somewhat arbitrary, especially given the small Antlia source population. To compare the luminous end of AMUSE-Antlia with AMUSE-Virgo and AMUSE-Field, we calculated the expected AGN detections in these samples for the stellar mass of AMUSE-Antlia and took the number of the XLF of the two other samples. The AMUSE-Field XLF gives $\sim 1.2$ and $\sim 0.6$ AGN in $\logLx=39.7 \textup{--} 40.5$ and $\logLx=40.5 \textup{--} 41.3$ bins.  AMUSE-Virgo XLF implies $\sim 0.2$ AGN in these same two bins.  As a result, one real Antlia AGN detection at the luminous end is between the expected $\sim 0.4$ AGN based on the AMUSE-Virgo XLF and $\sim 1.8$ AGN on the AMUSE-Field XLF, although the large uncertainty should be noted. For the less luminous bin of $\logLx=38.9 \textup{--} 39.7$, the number of Antlia AGN found relative to the total stellar mass is comparable with that of AMUSE-Field, and much higher than AMUSE-Virgo. Even though the luminous end of AMUSE-Antlia XLF is hard to constrain due to the small number of AGN, we conclude that the nuclear SMBH activity is higher in a dynamically active environment like the Antlia cluster and the Field, but lower in a more relaxed environment, like Virgo and Fornax.

\subsection{Occupation Fraction}\label{sec:occupation_fraction}

We compare the black hole activity in ETG of different samples through their AGN occupation fractions. Two Antlia sources hosted by LTG -- Antlia $88$ and Antlia $98$ are excluded. Unlike AMUSE-Virgo, AMUSE-Field, and Fornax, in which most galaxies have
$\sim 5\ \mathrm{ks}$ on-axis snapshots, galaxies in the three AMUSE-Antlia fields have been observed at a variety of detection limits. To ensure a fair comparison, we restrict our study to nuclear X-ray sources of $\logLx \geqslant 38.9$ in the energy band of $0.5$--$8\ \mathrm{keV}$. The luminosity threshold is chosen as the least luminous nuclear X-ray source of AMUSE-Antlia since it has the highest detection limit. The $0.3\textup{--}10\ \mathrm{keV}$ nuclear source luminosities of AMUSE-Virgo and AMUSE-Field are given in \citet{2010ApJ...714...25G} and \citet{2012ApJ...747...57M}, respectively. We convert them to $0.5$--$8\ \mathrm{keV}$ by assuming a single power-law with a photon index of $\Gamma=2$, as used in the two works. Furthermore, since galaxies in these four samples have distinct stellar mass functions, we categorize them into four equally sized stellar mass bins, which range from $ \logMs = 8 \textup{--}11.6$.
The occupation rate $f_\mathrm{occ}$ is defined as the number of AGN over the galaxy population. To calculate the uncertainties in the occupation rate, we adopt the posterior probability density function (PDF) based on Bayesian analysis \citep[Equation 2]{2022arXiv221013717S} ,

\begin{equation}\label{eq:post-pdf}
\begin{array}{l}
	 P(f_\mathrm{occ}|N_\mathrm{gal},N_\mathrm{AGN}  ) \propto \\  \int P(N_\mathrm{gal},N_\mathrm{AGN}| f_\mathrm{occ},\lambda ) P(\lambda )\mathrm{d}\lambda \ ,
\end{array}
\end{equation}

where $N_\mathrm{gal}$  is the number of galaxies. $N_\mathrm{AGN}$ is the detected number of AGN and $\lambda$ stands for its expectation value.  
On the right side of Equation \ref{eq:post-pdf}, the joint likelihood can be written as two parts,

\begin{equation}\label{eq:joint-likelihood}
	P(N_\mathrm{gal},N_\mathrm{AGN}| f_\mathrm{occ},\lambda ) = P(N_\mathrm{gal})P(N_\mathrm{AGN}|\lambda) \ .
\end{equation}

The number of galaxies $N_\mathrm{gal}$ is known,  which gives $P(N_\mathrm{gal})=1$.  Now, $\lambda = f_\mathrm{occ}  N_\mathrm{gal}$, and $P(\lambda)=\delta(\lambda - f_\mathrm{occ}N_\mathrm{gal})$, where $\delta$ is the Dirac function, so $f_\mathrm{occ}$ can be omitted from the expression. Then, Equation \ref{eq:post-pdf} reduces to
\begin{equation}
    P(f_\mathrm{occ}|N_\mathrm{gal},N_\mathrm{AGN}  )\propto P(N_\mathrm{AGN}|\lambda)\ .
\end{equation}

$P(N_\mathrm{AGN}|\lambda)$ follows the binomial distribution, where, 
\begin{equation}
\begin{array}{l}
  P(N_\mathrm{AGN}|\lambda)=\\ \frac{N_\mathrm{gal}!}{N_\mathrm{AGN}!(N_\mathrm{gal} - N_\mathrm{AGN})!} f_\mathrm{occ}^{N_\mathrm{AGN}}(1-f_\mathrm{occ})^{N_\mathrm{gal} - N_\mathrm{AGN}}\ .  
  \end{array}
\end{equation}

Thus, we can get the PDF of the occupation rate in order to calculate uncertainties.

\begin{deluxetable}{ccccc}
\tablenum{2}
\tablecaption{ETG AGN and host galaxy population of different samples}
\tablewidth{0pt}
\tabletypesize{\footnotesize}
\tablehead{
\colhead{Sample} &  \colhead{Tiny $M_\star$} &\colhead{Small $M_\star$} &  \colhead{Medium $M_\star$} & \colhead{Large $M_\star$}\\
\cline{2-2}
\cline{3-3}
\cline{4-4}
\cline{5-5}
\colhead{} &\colhead{ \scriptsize{$N_{\mathrm{AGN}}/N_{\mathrm{gal}}$} }&\colhead{ \scriptsize{$N_{\mathrm{AGN}}/N_{\mathrm{gal}}$} } &\colhead{ \scriptsize{$N_{\mathrm{AGN}}/N_{\mathrm{gal}}$} } &\colhead{ \scriptsize{$N_{\mathrm{AGN}}/N_{\mathrm{gal}}$} } }
\decimalcolnumbers
\startdata
Antlia & $0/25\ (0\%)$ & $1/16\ (6.3\%)$ & $3/8\ (37.5\%)$ & $3/5\ (60\%)$ \\
Field & $1/23\ (4.3\%)$ & $0/17\ (0\%)$ & $8/21\ (38.1\%)$ & $20/30\ (66.7\%)$\\
Virgo & $0/5\ (0\%)$ & $0/45\ (0\%)$ & $4/33\ (12.1\%)$ & $3/10\ (30\%)$ \\
Fornax & $0/0\ (0\%)$ & $0/8\ (0\%)$ & $3/14\ (21.4\%)$ & $0/5\ (0\%)$ 
\enddata
\tablecomments{($ 1 $) Sample Name.  (2)\textup{--}(5) Number of ETG AGN with $\logLx\geqslant38.9$ in $0.5$--$8\ \mathrm{keV}$ energy band, number of ETG and its occupation fraction in four stellar mass bins.  Tiny $M_\star$, small $M_\star$, medium $M_\star$, and large $M_\star$ correspond to the logarithmic galaxy stellar masses of $8\textup{--}8.9$, $8.9\textup{--}9.8$, $9.8\textup{--}10.7$, $10.7\textup{--}11.6$, as shown in Figure \ref{fig:OccFraction}.}
\label{table:occupation-rate}
\end{deluxetable}

\begin{figure}[ht!]
\epsscale{1.2}
\plotone{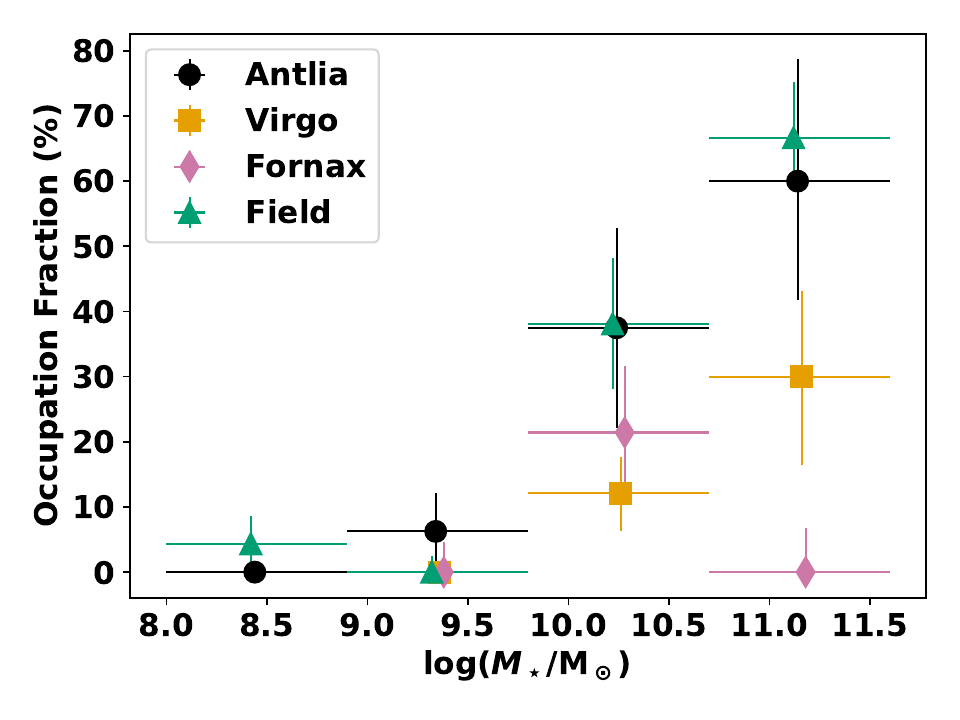}
\caption{The occupation fraction of AMUSE-Antlia,  AMUSE-Field, AMUSE-Virgo, and Fornax in four mass bins. The data are restricted to nuclear X-ray sources with $\logLx \geqslant 38.9$ in the $0.5\textup{--}8\ \mathrm{keV}$ band and logarithmic galaxy stellar masses of $8\textup{--}8.9$, $8.9\textup{--}9.8$, $9.8\textup{--}10.7$, and $10.7\textup{--}11.6$. The occupation fractions of Antlia and Field samples are higher than Virgo and Fornax.}
\label{fig:OccFraction}
\end{figure}

The occupation rate is likely biased high for AMUSE-Field and that of AMUSE-Antlia may have been underestimated due to the Eddington incompleteness. When restricted to the same X-ray luminosity limit of $\logLx \geqslant 38.9$, and the same stellar mass range of $10 \leqslant\log (M_\star/\mathrm{M_\odot})<11.6$, we find that the ETG AGN occupation fraction is $55_{-14}^{+13}\%$ in Antlia. This fraction is $55_{-7}^{+7}\%$ for the Field, $18_{-6}^{+6}\%$ for Virgo, and $17_{-10}^{+11}\%$ for Fornax. This finding indicates that the black hole activity is enhanced in Antlia, relative to Virgo or Fornax. Also, we compare their occupation fractions in four mass bins in Figure \ref{fig:OccFraction}. The number of AGN with $\logLx\geqslant38.9$ in $0.5$--$8\ \mathrm{keV}$ energy band and the number of galaxies in each bin are listed in Table  \ref{table:occupation-rate}. We do not include the lowest mass bin for Fornax and Virgo due to their very small numbers of galaxies. In the two massive bins, the occupation fractions of Antlia and Field are higher than Virgo and Fornax.

\begin{figure}[ht!]
\epsscale{1.2}
\plotone{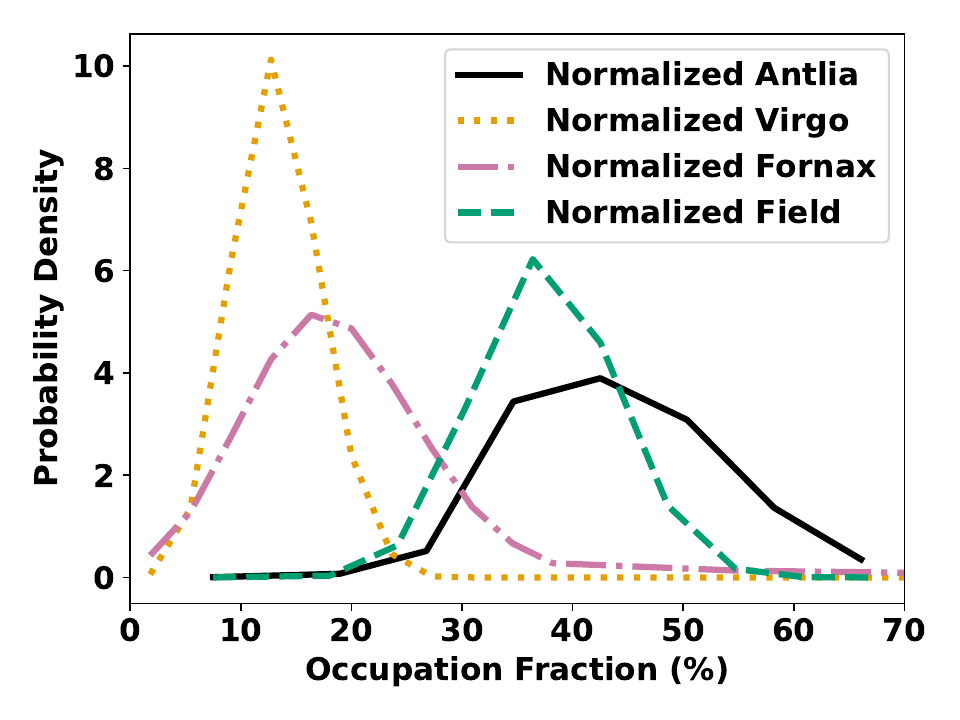}
\caption{The probability density functions of the weighted bootstrap replications. Note that the Field refers to the AMUSE-Field sample. As a result, the outcome occupation fraction of the AMUSE-Antlia, AMUSE-Field, AMUSE-Virgo, and Fornax samples are $45\% \pm 10\%$, $38\% \pm 6\%$,  $13\% \pm 4\%$, and $20\% \pm 8\%$, respectively.  In conclusion, the AGN activity is similar in AMUSE-Antlia and AMUSE-Field samples, but both are higher than AMUSE-Virgo and Fornax.}
\label{fig:WB-newfOcc}
\end{figure}

To fully normalize the impact of different $M_\star$ distributions, we use a weighted bootstrap method, which gives a posterior probability for each item when bootstrapping. We apply the same method on AMUSE-Antlia, AMUSE-Field,  AMUSE-Virgo, and Fornax, setting the GAMA-II survey ETG stellar mass function \citep{2016MNRAS.457.1308M} as the standard. The technique details are described in Appendix \ref{appendix:WeightedBootstrap}. As a result, the new occupation fractions of the normalized samples are $45\%\pm10\%$, $38\%\pm6\%$, $13\%\pm4\%$, and $20\%\pm8\%$ for AMUSE-Antlia, AMUSE-Field, AMUSE-Virgo, and Fornax, respectively, as shown in Figure \ref{fig:WB-newfOcc}. In conclusion, the AGN activity of Antlia and Field are similar, while both are higher than Virgo and Fornax. 

\section{Discussion}\label{sec:discussion}

We find that the black hole activity of AMUSE-Antlia is similar to that of AMUSE-Field, and higher than AMUSE-Virgo and Fornax. Note that the AMUSE-Field includes systems from various different environments. Among the AMUSE-Field galaxies with known group membership status, $78\%$ are group galaxies, while $22\%$ are non-group members. However, the black hole activity for the group and non-group members are nearly identical \citep[Section 5]{2012ApJ...747...57M}. Overall, we consider the AMUSE-Field sample represents a non-cluster environment. In this context, an intriguing question arises: what is responsible for the enhanced black hole activity in the Antlia cluster, which exceeds the other two clusters and reaches the activity level of non-cluster galaxies?

We find 5 AGN (3 ETG AGN and 2 LTG AGN) in the NGC 3258 field, which exceeds the 3 AGN found in the NGC 3268 field and 1 in the southeast field. \citet{2015MNRAS.452.1617H} note that the NGC 3258 subcluster is the younger structure in the Antlia cluster, based on a larger velocity dispersion of galaxies around NGC 3258 and abundant \hi \ gas content. 
\citet{1997ApJ...485L..17P} also conclude that the intragroup gas of NGC~3258 has low metalicity based on the study of its X-ray halo.

The AGN activity is enhanced in Antlia, compared to Virgo and Fornax, while within Antlia, the youngest subcluster has the highest AGN activity. 
A picture starts to emerge that a dynamically young environment is responsible for triggering AGN.

Cold gas can fuel AGN accretion. Thus, it is natural to link the cold gas content with AGN activity.  Antlia is found to retain a large population of gas-rich galaxies.
However, these gas-rich galaxies in Antlia are not strongly linked with AGN.  \citet{2015MNRAS.452.1617H} present a $4.4\ \mathrm{deg^2}$ \hi\ mosaic survey that fully covers the three AMUSE-Antlia fields with the Karoo Array Telescope (KAT-$7$). Four LTGs within the AMUSE-Antlia footprint (Antlia $93$, Antlia $98$, Antlia $120$, and Antlia $212$) have \hi\ detections. Only the BCD, Antlia $98$, contains a nuclear X-ray source (see Section \ref{ssec:bcd_xrc} for details).  
In addition, \citet{2019ApJ...882..132C} use the Atacama Pathfinder Experiment telescope (APEX) to study the CO $(2\textup{--}1)$ content as an $\mathrm{H_2}$ tracer of $72$ Antlia galaxies, $20$ of which are in the AMUSE-Antlia fields. As a result, four ETG (Antlia $72$, Antlia $111$, Antlia $222$, and Antlia $224$) have CO $(2\textup{--}1)$  detections. However, none of them is paired with a nuclear X-ray source. Also, $7$ out of the $9$ nuclear X-ray source hosting galaxies in AMUSE-Antlia have APEX observations, but none of them is confirmed to contain CO. 
The non-detection of cold gas is not a result of depletion by black hole accretion. Taking the maximum Eddington ratio of $10^{-4}$ as suggested in \citet{2012ApJ...747...57M} for the low-luminosity AGN (LLAGN) in the AMUSE-Virgo and AMUSE-Field surveys \citep[see the left panel of Figure 5 in][]{2012ApJ...747...57M}, the accretion rate of an AGN is expected to be $ \sim 10^{-4} \ \mathrm{M_\odot \ yr^{-1}}$. Antlia contains \hi \ gas and molecular gas with $M_{\text{\mbox{H\,{\sc i}}}} \sim 1.5\times10^{10}\ \mathrm{M_\odot}$ \citep[Table 2]{2015MNRAS.452.1617H} and $M_\mathrm{mol} \sim 9\times10^9\ \mathrm{M_\odot}$ \citep[Table 2]{2019ApJ...882..132C}, and the depletion timescale would be $10^{13}\ \mathrm{yr}$, greatly exceeding the Hubble time. 
The lack of correlation between cold gas content and AGN activity may be due to the limited detection sensitivity.

It is also possible to maintain LLAGN without cold gas.  A promising mechanism is the radiatively inefficient accretion flow (RIAF), such as the advection-dominated accretion flow (ADAF) model (e.g. \citealt{2014ARA&A..52..529Y}; \citealt[chap. 4.4]{2013peag.book.....N}). In this model, the accretion flow can be very hot, even reaching the virial temperature. ADAFs inefficiently convert the gravitational energy to radiative electromagnetic energy because much of the energy is advected into the black hole, thus producing LLAGN. 
In this scenario, 
nuclear X-ray sources are not necessarily expected to be associated with cold gas. 

The enhanced AGN activity may be related to the role of ram pressure stripping.
Ram pressure stripping can induce a loss of angular momentum of the gas, causing gas, cold and hot, to flow towards the center and trigger AGN. For example, \citet{2017Natur.548..304P} study seven jellyfish galaxies with clear ram pressure stripping morphology and find 6 of them host AGN.
The infall of the NGC 3258 subcluster is likely to have caused ram pressure stripping. There is also clear evidence for ram stripping in Fornax \citep[e.g.][]{2023A&A...673A.146S,2017ApJ...835...19S,2017ApJ...834...74S,1997ApJ...482..143J} and Virgo \citep[e.g.][]{2014A&A...564A..67B,2022A&A...667A..76J,2019AJ....158....6S,1979ApJ...234L..27F}. 
However, these two clusters may be experiencing different stages of ram pressure stripping compared to Antlia. All three clusters contain molecular gas (Virgo: e.g. \citealt{1989ApJ...344..171K}, Fornax: \citealt{2021A&A...648A..32K}), while only Antlia has sufficient \hi \ gas and both Virgo \citep[e.g.][]{1989ApJ...344..171K,2007A&A...474..851D,2010MNRAS.409..500O} and Fornax \citep{2021A&A...648A..31L} are \hi \ deficient. \citet{2014A&A...564A..67B} suggest that the molecular gas is not stripped as efficiently as the atomic gas. 
Therefore, Antlia is likely to be in an early stage of ram pressure stripping, in which its gaseous supply is yet completely removed.

\section{Summary and Conclusions}\label{sec:conclusion}

In this work, we present a study of the nuclear X-ray sources of member galaxies in the Antlia cluster using deep {\it Chandra} observations. 
We also include optical data from CTIO and radio observations from MeerKAT.
The detection rate of nuclear X-ray sources is $7/84\ (8.3\%)$ for ETG and $2/8\ (25\%)$ for LTG.    All nuclear X-ray sources, but one, have radio counterparts from broad bandwidth MeerKAT L-band observation. These sources in ETG, which typically lack star formation, are considered to be AGN. 
According to the fundamental plane (Appendix \ref{appendix:FundPlane}), the estimated black hole masses range from $3\times10^5\ \mathrm{M_\odot}$ to $5\times10^7\ \mathrm{M_\odot}$, with an error of $0.3\textup{--}0.4$ dex.
We perform a stacking analysis for galaxies in which nuclear X-ray  sources are not individually detected, yielding a detection of $L_\mathrm{0.5\textup{--}8\ keV} = (8.0\pm2.6)\times10^{37}\ \mathrm{erg\ s^{-1}}$ with a SNR $3.1$, which implies the existence of low luminosity nuclear X-ray activity below our detection limit. 
For low mass galaxies with $\log (M_\star/\mathrm{M_\odot}) < 8.5$ , we obtain a $3\sigma$ upper limit on their nuclear X-ray luminosity of $L_\mathrm{0.5\textup{--}8\ keV} =  3.3\times10^{37}\ \mathrm{erg\ s^{-1}}$.

The Antlia cluster, as a non-cool core cluster with an ongoing merger, presents a typical dynamically young environment, in sharp contrast with the relatively relaxed, cool core clusters, Virgo and Fornax. 
These three nearest clusters provide an ideal laboratory for studying the environmental effect on the black hole activity.  When restricted to the same X-ray luminosity limit of $\logLx \geqslant 38.9$, and the same stellar mass range of $10 \leqslant\log (M_\star/\mathrm{M_\odot})<11.6$, we find that the ETG AGN occupation fraction is $55_{-14}^{+13}\%$ in Antlia. This fraction is $55_{-7}^{+7}\%$ for the Field, $18_{-6}^{+6}\%$ for Virgo, and $17_{-10}^{+11}\%$ for Fornax. This finding indicates that the black hole activity is enhanced in Antlia, relative to Virgo or Fornax, consistent with the study of their AGN XLF.

An early stage of ram pressure stripping may be responsible for the enhanced AGN activities in a dynamically young environment, such as the non-cool core cluster, Antlia, particularly its young subcluster NGC~3258. 
There is more cold gas in Antlia member galaxies, especially in the young NGC 3258 subcluster, than other clusters, but we do not find a direct link between the detected cold gas and the AGN-hosting ETG, which may be due to the limited detection sensitivity. 
Meanwhile, the LLAGN may be maintained by the accretion of hot gas.

\begin{acknowledgments}

The authors thank the anonymous reviewer for their helpful comments on the manuscript.

Z.H. and Z.L. acknowledge the support of the National Natural Science Foundation of China (grant 12225302).

Y.S. acknowledges support from Chandra X-ray Observatory
grants GO1-22104X, GO2-23120X and NASA Grants 80NSSC22K0856. 

K.M.H. acknowledges financial support from the grant CEX2021-001131-S funded by MCIN/AEI/ 10.13039/501100011033, from the coordination of the participation in SKA-SPAIN, funded by the Ministry of Science and Innovation (MCIN).

W.F. and C.J. acknowledge support from the
Smithsonian Institution, the {\it Chandra} High Resolution
Camera Project through NASA contract NAS8-03060,
and NASA Grants 80NSSC19K0116, GO1-22132X, and
GO9-20109X.

A.S. acknowledges support through a Clay Fellowship.

Z.H. acknowledges Fangzheng Shi and Tao Wang for helpful discussion on mathematics and ETG mass function.

This research made use of \texttt{photutils}, an \texttt{astropy} package for
detection and photometry of astronomical sources \citep{larry_bradley_2021_5796924}.
\end{acknowledgments}

\begin{appendix}

\section{The Fundamental Plane}\label{appendix:FundPlane}
The fundamental plane indicates the correlation among radio flux, X-ray flux, and black hole mass, as shown in \cite[Equation 8]{2019ApJ...871...80G}: 

\begin{equation}\label{eq:fundamental-plane}
    \mu = 0.55\pm0.22 + (1.09\pm0.1) R + (-0.59^{+0.16}_{-0.15})X,
\end{equation}

where $\mu = \log (M_\mathrm{BH} / 10^8\ \mathrm{M_\odot}) $, $R = \log(L_\mathrm{R}/10^{38}\ \mathrm{erg\ s^{-1}})$ at $5\ \mathrm{GHz}$, and $X = \log (L_\mathrm{X}/10^{40}\ \mathrm{erg\ s^{-1}})$ in $2-10\ \mathrm{keV}$. 

\begin{figure}[ht!]
\epsscale{1.2}
\plotone{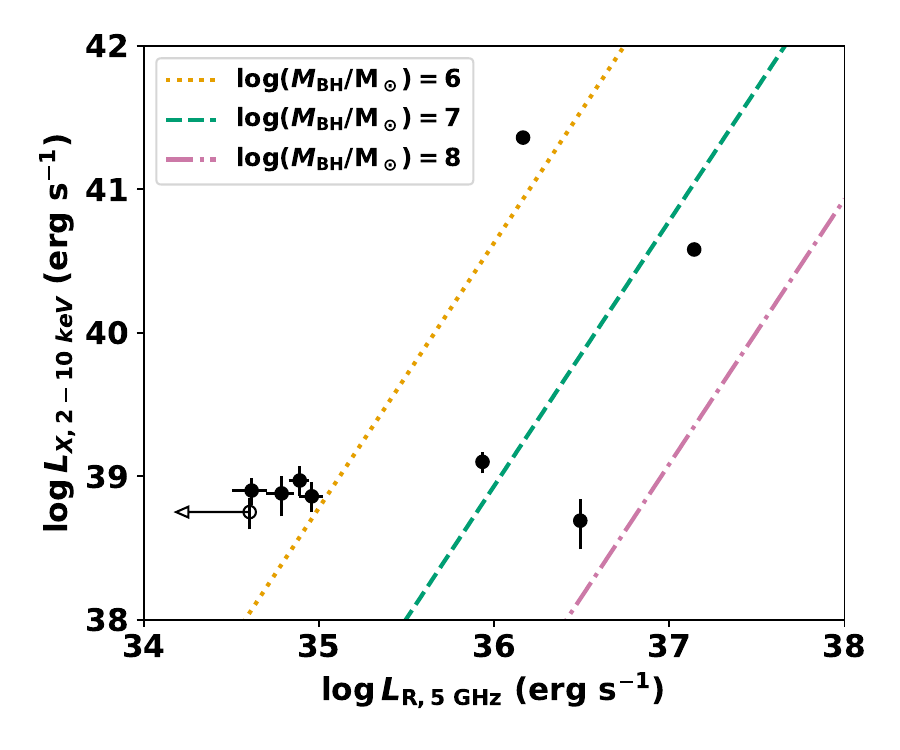}
\caption{The fundamental plane correlation of AGN mass, $5\ \mathrm{GHz}$ 
radio luminosity, and $2\textup{--}10\ \mathrm{keV}$ hard X-ray luminosity. The $9$ Antlia galaxies hosting nuclear X-ray sources are shown as the black dots with $1\sigma$ error. Antlia $105$ does not have a significant radio source, so here we use $3$ times the root-mean-square background level as the radio term to calculate the black hole mass and present it with an open circle. The yellow dotted, green dashed, and purple dash-dotted line indicate black hole mass $\log (M_\mathrm{BH} / \mathrm{M_\odot})$  from $6$ to $8$. As a result, if the galaxies do host AGN, the black hole masses fall in a reasonable range $ 6\lesssim \log (M_\mathrm{BH} / \mathrm{M_\odot}) \lesssim 8$.\label{fig:LXvsLR}}
\end{figure}

We measure the $1.28 \ \mathrm{GHz}$ radio source flux by fitting each source image with a $2$-dimensional Gaussian distribution using Python package {\tt photutils} \citep{larry_bradley_2021_5796924}.  We convert the flux to $5\ \mathrm{GHz}$ assuming a power-law spectrum $S\propto \nu^{-0.7}$ \citep{2002AJ....124..675C}, where $S$ is the measured flux and $\nu$ is the frequency. As shown in Figure \ref{fig:LXvsLR}, the $9$ galaxies with nuclear X-ray sources and radio emission have black hole mass $ 6\lesssim \log (M_\mathrm{BH} / \mathrm{M_\odot}) \lesssim 8$, which is typical for AGN. 

\section{The Weighted Bootstrap Method}\label{appendix:WeightedBootstrap}

Bootstrap is a kind of Bayesian method assuming a static posterior probability of $1/n$ for each individual value, where $n$ is the size of the sample \citep{1979...Efron...Bootstrap, 1981...Rubin...BayesianBootstrap}.  
The AGN occupation fraction is defined as the ratio of AGN numbers to the galaxy population. For the purpose of comparing the AGN occupation fractions of samples with different stellar mass distributions, we set the bootstrap posterior probability to normalize the replicated sample to the local spheroidal stellar mass distribution \citep{2016MNRAS.457.1308M}. We refer to this method as the weighted bootstrap method.

\textbf{1. A mathematical description}. Suppose that we have an observed sample with size $n$, let $\vect{x} = (x_1,x_2,\cdots,x_n)$ and $x_i$ denotes the $i$th item. Consider a set of bins $\vect{b}=(b_1,b_2,\cdots,b_m)$ with size $m$, where $\vect{b}$ is chosen to ensure that each bin corresponds to at least one $x_i$ and each $x_i$ falls in bin $b_j$, say: $b_j-\frac{1}{2}\Delta b\leqslant x_i < b_j+\frac{1}{2}\Delta b$, where $\Delta b$ is the bin width.  A statistic $\hat{f}$ estimates a parameter $f$ based on a distribution  $\Phi(x)$. To normalize $\vect{x}$ to $\Phi(x)$, so that the statistic $\hat{f}$ can be applied to $\vect{x}$, a posterior probability $P_i$ for each $x_i$ is given as,

\begin{equation}\label{eq:WeightedBootstrapPosteriorProbability}
\left\{
\begin{array}{l}
P_i\propto \Phi(b_j)/N_j \\
\sum_{i=1}^{n}P_i = 1,
\end{array}\right.
\end{equation}

where $N_j$ is the number of $\vect{x}$ elements that fall in $b_j$. Thus, a posterior probability set $\vect{P} = (P_1,P_2,\cdots,P_n)$ 
is constructed,
corresponding to $\vect{x}$. 

A weighted bootstrap replication generates a random sample of size $n$ from $\vect{x}$ with a weight factor in $\vect{P}$. Applying $\hat{f}$ to one replicated sample gives one estimate of parameter $f$. After many replications, the distribution of the replicated items approaches $\Phi(x)$.  Finally, the result of $\hat{f}$ on $\vect{x}$ is calculated from all possible bootstrap estimations of  $f$.

\textbf{2. Realization.} \ We aim to compare the AGN occupation fractions of three samples, AMUSE-Virgo, AMUSE-Field, and AMUSE-Antlia. 
We implement this weighted bootstrap method to correct for their different stellar mass distributions (see right panel of Figure \ref{fig:AGN_XLF})
which can strongly bias the occupation fraction.

Before the statistical procedure, we first restrict the stellar mass to the range of $9\leqslant \log\ (M_\star / \mathrm{M_\odot})<11.7$. Also, we keep AGNs with $0.5\textup{--}8\ \mathrm{keV}$ X-ray luminosity $\logLx \geqslant 38.9 $, which is the median detection limit of AMUSE-Antlia, the highest among the four samples. With these constraints, the occupation fractions of AMUSE-Antlia, AMUSE-Field, AMUSE-Virgo, and Fornax are $26^{+8}_{-8}\%$,  $42^{+6}_{-6}\%$, $12^{+3}_{-4}\%$, and $11^{+5}_{-6}\%$, respectively.

The Schechter function \citep{1976ApJ...203..297S} describes the probability density function for each galaxy in mass space. 

\begin{equation} \small
\begin{array}{ll}
     \Phi(\log M)\mathrm{d} \log M = 
      &  \ln(10)\phi^{*}10^{\log(M/M^{*})(\alpha+1)} \\
 &  \exp(-10^{ \log (M/M^{*})})\mathrm{d} \log M,
\end{array}
\end{equation}

where $\phi^*$ is the normalization constant, $M^*$ is the characteristic mass of the `knee' in the mass function, and $\alpha$ is the slope at the low mass end. We adopt the parameters $\log(M^*/\mathrm{M_\odot})=10.74\pm0.026$, $\alpha=-0.525\pm0.029$ and $\phi^*=3670\pm200\ \mathrm{dex^{-1}\ Mpc^{-3}}$ of the local spheroidal stellar mass distribution, according to the Galaxy and Mass Assembly survey phase two (GAMA-II) \citep{2016MNRAS.457.1308M}. 

\begin{figure}[ht!]
\epsscale{1.1}
\plotone{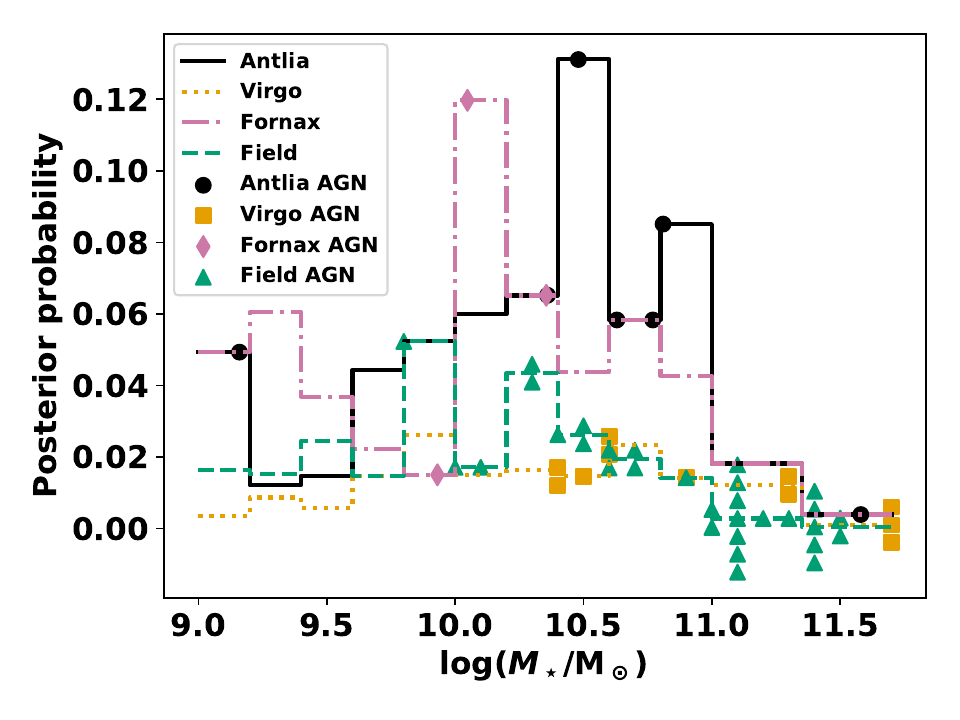}
\caption{The posterior probability $P_i$ of the weighted bootstrap method according to Equation \ref{eq:WeightedBootstrapPosteriorProbability}. The sum of the posterior probability is less than $1$ because there are multiple galaxies in one bin. Each marker represents a galaxy hosting an AGN and its corresponding stellar mass.  The  $5$  AMUSE-Antlia AGN with host stellar masses around $10.7\ \mathrm{dex}$ have higher posterior probabilities, which contributes to the increase of the resulting occupation fraction. }
\label{fig:WeightedBootstrap}
\end{figure}

We then calculate the occupation fraction of AMUSE-Antlia by normalizing the stellar mass distribution to the Schechter function with the weighted bootstrap method. For convenience, $\vect{x}$ stands for the observed AMUSE-Antlia sample with $n$ galaxies, and $x_i$ indicates the stellar mass of the $i$th member galaxy. There are $12$ stellar mass bins in $\vect{b}$, the first $10$ bins have bin widths of $0.2 \ \mathrm{dex}$, while the last two have widths of $0.3\ \mathrm{dex}$. We generate weighted bootstrap replications with a posterior probability $P_i$, according to Equation \ref{eq:WeightedBootstrapPosteriorProbability}, as plotted in Figure \ref{fig:WeightedBootstrap}. After $100,000$ replications, the mean stellar distribution fits $\Phi(x)$ well. The statistic $\hat{f}$ calculating the occupation fraction is applied to each replication.  Finally, the weighted bootstrapped distribution of $\hat{f}$ on $\vect{x}$ is shown in Figure \ref{fig:WB-newfOcc}.

The same process is also applied to AMUSE-Field, AMUSE-Virgo, and Fornax. As a result, the new occupation fractions of the normalized samples are  $45\% \pm 10\%$,  $38\% \pm 6\%$, $13\% \pm 4\%$, and $20\% \pm 8\%$ for normalized Antlia, Field, Virgo, and Fornax, respectively. Here, the bootstrapped occupation fraction of AMUSE-Antlia is much higher than the original value. This is due to $5$ of $7$ AMUSE-Antlia AGN being hosted by ETG with stellar masses around $10.7\ \mathrm{dex}$, which is the `knee', the peak of Schechter function. Due to their higher posterior probability, these five items significantly contributed to the substantial increase in the outcome. Similarly, \citet{2019ApJ...874...77L} normalize the stellar mass distributions of AMUSE-Virgo and AMUSE-Field samples to Fornax, and find that Virgo and Fornax have similar AGN activity, both lower than AMUSE-Field, consistent with our findings.  In conclusion, the AGN activity of Antlia and AMUSE-Field are quite similar, both of which are much higher than the AGN activity in Virgo and Fornax.

\end{appendix}

\newpage

\bibliography{AN_AGN_manuscript.bib}{}
\bibliographystyle{aasjournal}

\end{document}